\documentclass[useAMS,usenatbib,onecolumn]{mn2e}
\usepackage{graphicx}
\usepackage{amssymb}
\usepackage{epstopdf}
\usepackage{natbib}
\usepackage{bm}
\usepackage{color}
\usepackage{times}
\usepackage{deluxetable}
\usepackage{ulem}

\def\xmm{{\sl XMM-Newton}}
\def\chandra{{\sl Chandra}}
\def\fuse{{\sl FUSE}}

\def\kmps{\hbox{km $\rm{s^{-1}}$}}

\def\feii{Fe~{\sc ii}}

\def\ni{N~{\sc i}}
\def\niv{N~{\sc iv}}
\def\nv{N~{\sc v}}

\def\cii{C~{\sc ii}}
\def\civ{C~{\sc iv}}

\def\ovi{O~{\sc vi}}

\def\oiv{O~{\sc iv}}
\def\oiii{O~{\sc iii}}
\def\oi{O~{\sc i}}

\def\siii{Si~{\sc ii}}
\def\siiii{Si~{\sc iii}}
\def\siiv{Si~{\sc iv}}

\def\sulii{S~{\sc ii}}

\def\alii{Al~{\sc ii}}
\def\pii{P~{\sc ii}}

\def\heii{He~{\sc ii}}

\newcommand\aj{AJ}%
\newcommand\apj{ApJ}%
\newcommand\apjs{ApJS}%
\newcommand\apss{Ap\&SS}%
\newcommand\aap{A\&A}%
\newcommand\mnras{MNRAS}%
\newcommand\ssr{Space~Sci.~Rev.}%

\title[The UV Absorption in Mrk 290]{Modeling Warm Absorption in HST/COS Spectrum of Mrk 290 with XSTAR}

\author[Zhang et al.]{S. N. Zhang$^{1,2,3,4}$\thanks{E-mail:
snzhang@pmo.ac.cn}, L. Ji$^{1,2,3}$, T. R. Kallman$^5$, Y. S. Yao$^6$, C. S. Froning$^7$,
\newauthor 
Q. S. Gu$^{3,4,8}$, G. A. Kriss$^9$\\
$^1$Purple Mountain Observatory, Chinese Academy of Sciences, Nanjing, 210008, China\\
$^2$Key Laboratory of Dark Matter and Space Astronomy, PMO, CAS, Nanjing, 210008, China\\
$^3$Collaborative Innovation Center of Modern Astronomy and Space Exploration, Nanjing, 210093, China\\
$^4$Key Laboratory of Modern Astronomy and Astrophysics, Nanjing University, Nanjing, 210093, China\\
$^5$NASA/Goddard Space Flight Center, Greenbelt, MD 20771, USA\\
$^6$Eureka Scientific, Inc., 2452 Delmer Street Suite 100, Oakland, CA 94602, USA\\
$^7$Department of Astronomy, University of Texas at Austin, Stop C1400, Austin, TX 78712-1205, USA\\
$^8$School of Astronomy and Space Science, Nanjing University, Nanjing, 210008, China\\
$^9$Space Telescope Science Institute, 3700 San Martin Drive, Baltimore, MD 21218, USA}

\begin{document}

\date{Accepted 2014 December 05. Received 2014 December 05; in original form 2014 July 23}

\pagerange{\pageref{firstpage}--\pageref{lastpage}} \pubyear{2014}

\maketitle

\label{firstpage}

\begin{abstract}
We present a new method to model a HST/COS spectrum, aimed to analyze 
intrinsic UV absorption from the outflow of Mrk 290, a Seyfert I galaxy.
We use newly updated XSTAR to generate 
photoionization models for the intrinsic absorption from the AGN outflow, 
the line emission from the AGN broad and narrow line regions, 
and the local absorption from high velocity clouds and Galactic interstellar medium.
The combination of these physical models accurately fit the COS spectrum.
Three intrinsic absorbers outflowing with velocities $\sim$500 \kmps~are identified,
two of which are found directly from two velocity components of the \nv~and \civ~doublets, 
while the third is required by the extra absorption in the Ly$\alpha$.
Their outflow velocities, ionization states and column densities are consistent with the lowest and moderately ionization warm absorbers (WAs) in the X-ray domain found by \chandra~observations,
suggesting an one-to-one correspondence between the absorbing gas in the UV and X-ray bands.
The small turbulent velocities of the WAs ($v_{\rm turb}\lesssim100$ \kmps) support our previous argument from the X-ray study that the absorbers originate from the inner side of the torus due to thermal evaporation.
Given the covering fractions of $\sim$65\% for the three WAs, we deduce that the lengths and the thicknesses of the WAs are comparable, which indicates that the geometry of WAs are more likely clouds rather than flat and thin layers.
In addition, the modeling of the broad line emission suggests a higher covering fraction of clouds when they are very closer to the black hole.

\end{abstract}


\begin{keywords}
quasars: absorption lines  --- ultraviolet: galaxies ---  quasars: individual: Mrk 290
\end{keywords}
\clearpage

\section{Introduction}
Ionized outflowing gas from the central engines of Active Galactic Nuclei (AGN) has been studied for thirty years \citep[since][]{halpern84}.
In about 50\% of Seyfert I galaxies, outflowing gas is characterized by blue-shifted absorption lines in the UV band \citep{crenshaw99} and/or the X-ray band \citep{reynolds97, george98}.
The gas has a high covering fraction, a total mass exceeding $\sim10^3\,\rm M_\odot$, 
and is outflowing at a rate of $\gtrsim0.1\,\rm M_\odot\,yr^{-1}$ \citep{kriss04}.
Theoretically, it may have a significant impact on the evolution of the host galaxy, when its kinetic luminosity reaches 0.5\% of the bolometric luminosity of AGN \citep{hopkins10}.
A significant fraction of Seyfert I galaxies can reach this criterion \citep[e.g., $\gtrsim$3 in 10 nearby Seyfert I galaxies,][]{crenshaw12}.
However, a key question that complicates the determination of kinetic luminosity of outflowing gas
is its distance from the black hole (BH), and hence its geometry.
In only a few sources are the distance and the geometry well constrained through careful studies in recent years \citep{steenbrugge11}.

Ideally, it is possible to probe the geometric and physical properties of photoionized outflowing gas by combining the information from UV and X-ray observations \citep[e.g.,][]{costantini10}.
The X-ray absorbing gas contains a multitude of ions with up to $\sim$100 observable transitions in X-ray spectra \citep[e.g.,][]{kaspi02}.
The ionization states cover a large range and overlap with those of the UV absorbing gas.
The gas at very high ionization states can only be detected in X-ray spectra, while
UV band spectra have advantages in detecting absorption lines from gas at low ionization states and 
with low column densities.
However, the observations in the two bands do not merely offer the complementarity of a broader-band view
of the ionization structure of the plasma.
UV observations have significantly higher spectral resolution and so are able to supply better kinematic information 
by resolving complicated profiles of absorption lines.
Components with different velocities can be fully distinguished,
and their turbulent velocities can be estimated.
If reliable column densities for several ionic species are obtained simultaneously, 
elemental abundances can be estimated \citep[e.g.,][]{arav07}.
The geometric and physical properties of the outflowing gas can be inferred from the distributions of ionization state, column density, velocity and abundance.
Moreover, relative positions of the absorbing gas to different emitting components are easier to be obtained in the UV band.  Any variability in the UV and X-ray bands will also constrain on the locations.

It has been proven difficult to perform joint analyses of UV and X-ray spectra,
because it is hard to find an one-to-one relation of absorbers between the two bands. 
The reason is that the ionization level and kinematics of both the UV absorbers and 
X-ray warm absorbers (WAs) are sometimes difficult to measure, and there is often only partial
overlap in the conditions of the two sets of absorbers  \citep{crenshaw03}.
Using recent  high spectral-resolution telescopes and instruments, 
there have been numerous  multi-wavelength UV-X-ray campaigns.
These  provide much insight into the physical conditions in the absorbing outflows.
However, a broad understanding of these outflows has yet to emerge, and 
the relationship between the X-ray and UV absorbing gas is still not understood in many cases.
In most cases, UV absorbing gas is likely associated with the X-ray WAs, 
but frequently one X-ray WA may correspond to a blend of several UV absorbers
(e.g., NGC 3516, \citealt{kraemer02}; NGC 4593, \citealt{ebrero13}; 1H0419-577, \citealt{gesu13};
Mrk 279, \citealt{scott04}; Mrk 509, \citealt{kriss12}; MR 2251-178, \citealt{reeves13}).
In other cases, the absorbing gas may favor a continuous distribution of ionization states in a smooth flow, instead of discrete components
(e.g., NGC 5548, \citealt{steenbrugge05}; NGC 7469, \citealt{scott05}; IRAS, 13349+2438 \citealt{lee13}).
In other cases, variability  of the absorbers interfere with identification
(e.g., NGC 3783, \citealt[][a \& b]{gabel03}; NGC 4051, \citealt{collinge01}).
In particular cases, e.g. in NGC 4151, strong emission lines in the X-ray band contaminate the detection of WAs at lower ionization states \citep{kraemer05}.
In summary, absorbers are complicated in the both bands, and their information is not easy to be combined.

In this work, we focus on a nearby Seyfert I galaxy, Mrk 290, which shows an extremely simple absorption system in the UV band \citep{kriss02}.
Meanwhile, only three X-ray WAs were detected in its outflow using the \chandra~High Energy Transmission Grating Spectrometer (HETGS) in our previous paper \citep[][hereafter Z11]{zhang11}.
This is therefore a relatively clean case with advantages for the illustration of joint analysis.
Mrk 290 fortunately has simultaneous observations with \chandra~and \fuse~in 2003 (paper related to \fuse~observations is in preparation).
However, the exposure time of that \fuse~observation is relatively short.
In order to demonstrate how our new method helps the joint analysis in the two bands, 
we use HST Cosmic Origins Spectrograph (COS) observations in 2009 instead, 
which has a good spectral resolution and a higher signal-to-noise ratio.
The most likely origin of the WAs in Mrk 290 is from a slow torus wind (several 100 \kmps), based on constraints from the X-ray band.
Thus it is possible that the UV and X-ray observations reveal the same components even when the 
observations are separated by years.

The prevalent method in UV studies is to compare individual lines in velocity space.
Information on different aspects of each absorption line,
such as column density, ionization state, covering fraction, broadening, and saturation,
is collected and combined to infer physical quantities of the outflowing gas.
The process requires laborious work.
In this work, we demonstrate a novel and simple method to combine the information from the UV and X-ray bands.
We fit the UV spectra directly using synthetic spectral models generated by the XSTAR\footnote{http://heasarc.gsfc.nasa.gov/lheasoft/xstar/xstar.html} photoionization code, 
the same modeling technique as was used for the X-ray data.

The construction of this paper is as follows.
We describe the data reduction in Section 2, and the XSTAR code in Section 3.
In Section 4, we decompose the UV spectrum into different physical components, and then describe corresponding models for them.
The detailed modeling of the ionized absorbers is given in Section 5, and the discussion and exploration is given in Section 6.
We end with a summary of our conclusions in Section 7.

\section{Data Reduction}

The COS was installed on HST in May 2009.
It has two medium-resolution gratings, G130M and G160M, covering the wavelength range 1135-1795 \AA~with spectral resolutions of $R\equiv \lambda/\Delta \lambda$ from 16,000 to 21,000 \citep{osterman11}. 
Mrk 290 was observed with COS on Oct. 28, 2009, with a total exposure time of 8 ks. 
The COS observations and the calibration data were retrieved from the Multimission Archive (MAST\footnote{http://archive.stsci.edu}) and processed with the pipeline CALCOS (version 2.19.6). 
Spectra extracted from individual exposures were weighted by their signal-to-noise ratios and then co-added to form a final spectrum. 
By comparing ISM absorption lines from Complex C \citep{wakker06}, \citet{narayanan10} found the zero-point offset of the COS spectrum to be +17 \kmps, which is corrected for in this paper.

In this paper, we adopt a new technique for the analysis of the UV data.
For the first time, we fit the UV data from COS  using techniques which are customary for X-ray data. 
To provide the potential capability for a joint analysis of the UV and X-ray bands, we convert the UV spectrum into the style usually used in the X-ray band. 
We used the IDL tool {\small PINTofALE}\footnote{http://hea-www.harvard.edu/PINTofALE/} to simulate a Pulse Height Amplitude (PHA) file from the co-added COS spectrum and also to generate a corresponding response (RSP) file by convolving the line spread functions (LSF) of G130M and G160M.   
We fit the data using the Interactive Spectral Interpretation System \citep[ISIS version 1.6.2,][]{houck02}.
During the fit, a model spectrum is folded through the RSP file in order to generate synthetic data, 
which is then compared with the observed data.

\begin{figure*} 
   \centering
   \includegraphics[angle=0,width=\textwidth]{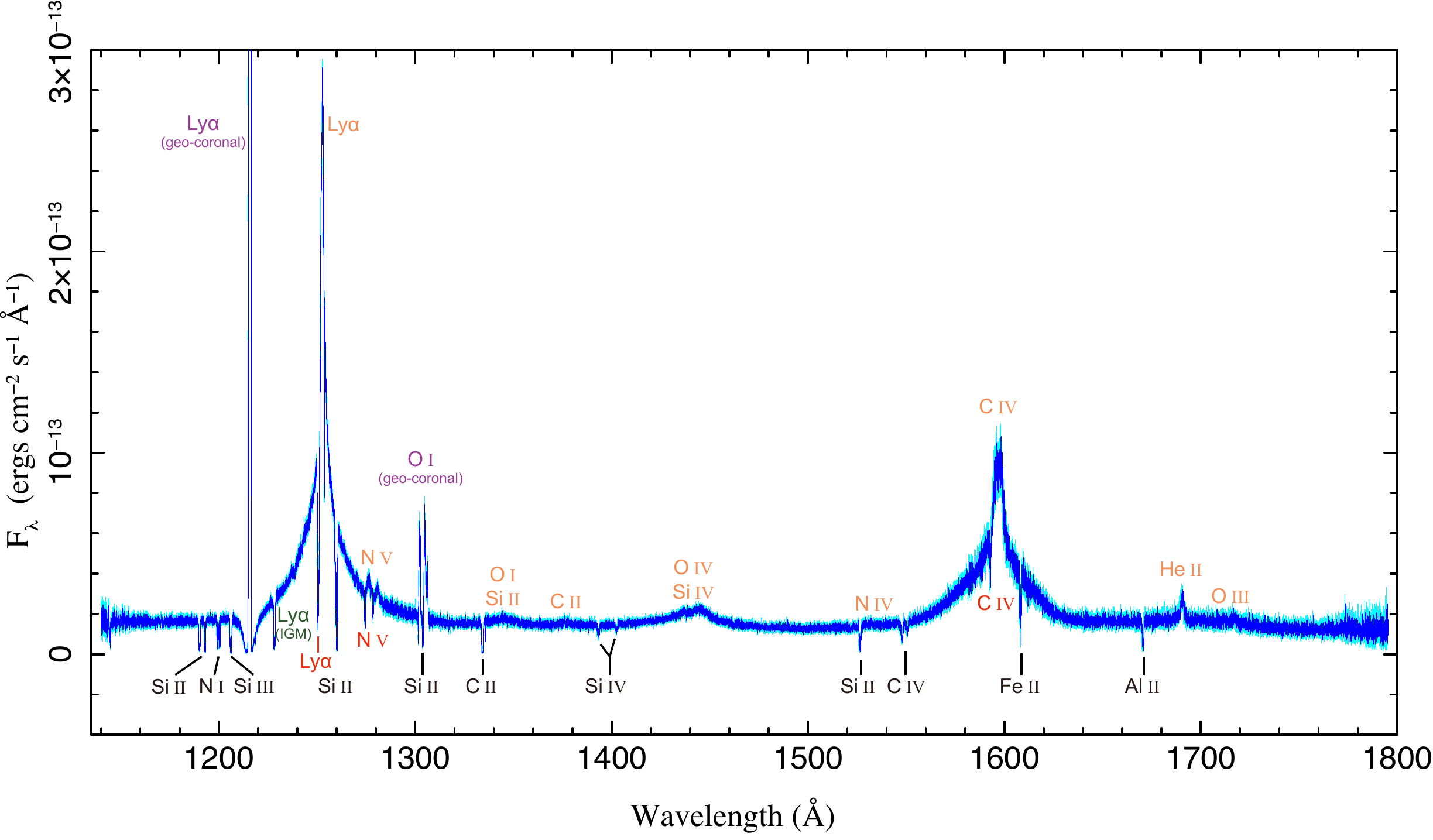} 
   \caption{The COS spectrum of Mrk 209 in the observer frame, where the data is in blue and the error bars are in cyan. The local absorption lines from HVCs and ISM are labeled with black notes, the prominent emission lines from the AGN with orange notes, and the intrinsic absorption lines from the outflow with red notes. The geo-coronal lines and IGM absorption line are also marked.}
   \label{fig:cos}
\end{figure*}

\section{XSTAR for UV band}

The models used to fit the data are calculated using the XSTAR photoionization code.  
XSTAR is used to compute physical conditions and synthetic spectra of photoionized plasmas. It performs the computationally intensive task of calculating profiles for a large number of lines, and allows interactive fitting of data to models via the use of tables of synthetic spectra.

XSTAR calculates the ionization balance,  atomic level populations and electron kinetic temperature for gas ranging from neutral atoms to fully stripped ions, and includes all elements with atomic number Z$\leq$30.    
The operation of XSTAR and its database have been described in \citet{kallman01} and \citet{bautista00}.    
XSTAR contains the microphysical rates and line lists needed to model optical, UV and X-ray emission from such gas.
Synthesizing spectra with the resolution of COS, however,  requires expanding many of the data structures used in the spectral calculation.   
This modification, contained in XSTAR version 2.2.1bn19 and newer, allows the synthesis of spectra with resolving power up to $\varepsilon/\Delta\varepsilon \simeq 8 \times 10^4$.  
The spectral resolution is an input parameter which can be controlled by the user, and the maximum value represents approximately an order of magnitude increase compared with previous versions.  
In addition, we have updated the line database to include several low ionization lines which are observed from the Milky Way and its halo.  
Although the increased resolving power results in a proportional increase in execution time, this is mitigated by the  availability of batch execution  using pvm\_xstar \citep{noble09}, so that useful grids of models can be computed within a few days.   
With these improvements, photoionization models generated by XSTAR cover the UV and X-ray bands simultaneously, allowing fitting results from the two bands to be compared with each other directly.    

In the case of XSTAR models, the free parameters used in fitting are usually the column density $N_{\rm H}$, the ionization parameter of the gas $\xi=4 \pi F/n$, (where $F$ is the incident flux in the 1-1000 Ry energy range, and $n$ is the hydrogen nucleus density), and the redshift $z$, while the temperature, density, turbulent velocity and elemental abundances are set by users, as shown in Table~\ref{tab:input}.
Beside these, a spectral energy distribution (SED) of the ionizing source is required by XSTAR for generating photoionization models.

\begin{deluxetable}{lccccccc}
\tablecolumns{8} 
\tablewidth{0pt} 
\tablecaption{The parameters for XSTAR table models}
\tablehead{
           \colhead{ }                                                    &
           \colhead{T }            				    &
           \colhead{$n$}         					    &
           \colhead{$b$\tablenotemark{a}}  				    &
           \colhead{$N_{\rm H}$}  					    &
           \colhead{Log($\xi$)}        				    &
           \colhead{Abundance\tablenotemark{b} }      &
           \colhead{$z$}                                         \\
           \colhead{  }                                                   &
           \colhead{(K)}                                                &
	  \colhead{($\rm{cm^{-3}}$)}                           &
           \colhead{(\kmps)}              			    &
           \colhead{($\rm{10^{20}\,cm^{-2}}$)}             &
           \colhead{($\rm{erg\,s\,cm^{-1}}$)}                &
           \colhead{     }                                                &
           \colhead{     }                  
                      }  
\startdata
\multicolumn{8}{l}{Local Absorption:} \\
HVC   & $5\times10^5$ & 0.2  & 25   &  1.12         & free & \citet{shull11} & free \\
ISM    & 8000                & 0.3  & 20   & free\tablenotemark{c}  & free  & solar & free  \\
    \hline\noalign{\smallskip}
    \multicolumn{8}{l}{AGN Radiation:} \\
BLR {\small I}  & 15000  & $10^{10}$ &  811    &  1000  & free  & solar & free \\
BLR {\small II}  & 15000  & $10^{10}$ &  3333  &  1000  & free  & solar & free \\
BLR {\small III}  & 15000  & $10^{10}$ &  6114  &  1000  & free  & solar & free \\
NLR       & 15000  & $10^3$     &  250    &  1000  & free  & solar & free \\
    \hline\noalign{\smallskip}
\multicolumn{8}{l}{Intrinsic Outflow Absorption:} \\
Absorber {\small I} & 6000 & $10^5$ & 16  & free & free & solar & free \\
Absorber {\small II} & 6000 & $10^5$ & 29  & free & free & solar & free \\
Absorber {\small III} & 30000 & $10^5$ & 115  & free & 2.2 & solar & free \\
\enddata
\tablenotetext{Note}{The XSTAR parameters include: temperature, hydrogen nucleus density, turbulent velocity, column density, ionization parameter, elemental abundances, and redshift.}
\tablenotetext{a}{The $b$ values are equal to the turbulent velocities $v_{\rm turb}$ for the local and intrinsic absorption lines. For the BLR and NLR, the values $b\approx \rm FWHM/1.665$ are used for producing required FWHMs of line profiles.}
\tablenotetext{b}{The solar abundances are defined by \citet{grevesse96}.}
\tablenotetext{c}{The column density of ISM has an upper limit as $0.64\times\rm{10^{20}\,cm^{-2}}$.}
\label{tab:input}
\end{deluxetable}

\section{Spectra Analysis for Individual Components}

In the COS spectrum, absorption and emission lines are heavily mixed (Fig.~\ref{fig:cos}).
To get the information of the intrinsic absorbers in Mrk 290, it is preferable to decompose all the physical components.
Three components need to be taken into account: AGN intrinsic emission, AGN intrinsic absorption, and local absorption. 
The other two components to notice are the Intergalactic Medium (IGM) absorption and the geo-coronal emission from the earth atmosphere.

Geo-coronal emission originates mainly from hydrogen and oxygen atoms in the exosphere of the Earth.
In this COS observation, there are three prominent geo-coronal features: Ly$\alpha$ (1215.7 \AA) and \oi~doublet (1302 \AA~and 1306 \AA).
During all the modeling, two narrow bands (1210--1225 \AA~\& 1300--1308 \AA) containing the geo-coronal lines are removed, though there is a local \siii~absorption line between the \oi~doublet.

Two nearby Ly$\alpha$ absorption lines from the IGM absorber 
are detected at 1228 \AA~and 1228.5 \AA, around the redshift $z=0.01028$,
and the corresponding \ovi~absorption is also identified in the \fuse~spectrum \citep{wakker09}.
This IGM absorber may be located in the outer halo of NGC 5987 \citep{narayanan10}. 
We model them simply with one negative Gaussian to reduce the $\chi^2$.

Eliminating the geo-coronal features and the IGM absorption, we discuss the three primary parts in detail as follows.
In this work, all the quoted errors refer to 90\% confidence level for one interesting parameter.

\subsection{Local Absorption}

The local absorption is from High Velocity Clouds (HVCs) and Galactic Interstellar Medium (ISM).
HVCs are low-metallicity clouds plunging toward our galaxy and interacting with its ambient halo gas \citep{collins07, tripp12}.
The line of sight toward Mrk 290 penetrates HVC Complex C, whose absorption lines in the COS spectrum have been studied by \citet{shull11} in detail.
Two absorbers were identified with blueshift velocities of -120 and -220 \kmps.
The absorber with -220 \kmps~velocity is at a higher ionization state, which shows shallow \siiii, \siiv~and \civ~absorption lines.
The absorber with -120 \kmps~velocity has both lower ionization lines (\oi, \ni, \cii, \sulii, \siii, \alii, \feii, \pii) 
and higher ionization lines (\siiii, \siiv, \civ, \nv).

Complex C is diffuse gas ionized by UV background photoionization and possible interactions with Galactic halo gas. 
We produce an XSTAR table model for its absorption, in which many input parameters are from the results of \citet{shull11}.
They obtained the column density of $N_{\rm H}=1.12\times10^{20}\,\rm{cm^{-2}}$, the hydrogen nucleus density of 0.2 cm$^{-3}$, and the turbulent velocity of $\sim$25 \kmps.
The metals have abundances less than 25\% solar metallicities \citep{grevesse96} as follows: C (10\%), O (10\%), N (1.5\%), S (17\%), Si (21\%), Fe (5\%), Al (4\%) and P (23.4\%), where the C abundance is adopted from the nearby lines of sight to Mrk 817 and Mrk 876.
We set the temperature as $5\times10^5$ K, which does not affect the table model much in the range of $10^5$ K \citep{slavin93} to $10^6$ K \citep{spitzer56}.
The ionization parameter and the redshift are left as free parameters in the model.

Galactic ISM absorption is attributed to ions, atoms, molecules and dust.
The molecular absorption is beyond the far UV end of the COS spectrum.
The neutral gas absorbs the Hydrogen Lyman series, i.e. Ly$\alpha$ in the COS spectrum.
But this narrow band (1210 - 1225 \AA) is removed in all the fittings due to the geo-coronal Ly$\alpha$.
Thus, the COS spectrum is mainly affected by the ions in warm ionized gas and by dust extinction.

The warm ISM gas is photoionized by the UV background radiation \citep{haardt12}.
In the COS spectrum, its absorption lines are similar to the lines from HVC.
We also use XSTAR to produce the photoionization table model for this component.
The density is set to 0.3 cm$^{-3}$, the temperature to 8000 K \citep{mckee77}, the turbulent velocity to 20 \kmps, 
and the metallicity abundance to solar values.
The column density, the ionization parameter and the redshift are left as free parameters.
Along the line of sight to Mrk 290, there is a relatively low Galactic absorption
$N_{\rm H}=1.76\times10^{20}\,\mathrm{cm}^{-2}$ obtained by 21 cm measurements \citep{kalberla05}.
Removing the contribution from HVC, we have an upper limit for the column density of the warm ionized gas in ISM of $N_{\rm H}=0.64\times10^{20}\,\mathrm{cm}^{-2}$.

Galactic dust extinction attenuates UV continuum  radiation, which can be corrected using extinction curves derived from other lines of sight.
We use the extinction curve formula described by \citet{gordon09} that covers the wavelength range from 910 \AA~to 3.3 $\mu$m.
The parameters used are a total-to-selective extinction ratio of $R_{\rm v}=A({\rm V})/E({\rm B-V})=3.1$ \citep{cardelli89}, and an extinction of $E({\rm B-V})=0.012$ at the line of sight to Mrk 290 \citep{schlafly11}.
There is no free parameter for the dust extinction component.

\subsection{Intrinsic Radiation from Mrk 290}

It is generally assumed that AGN intrinsic emission in the UV band comes from the accretion disk, broad line region (BLR) and narrow line region (NLR).
To fit the entire spectrum, it is possible to use phenomenological spline fits \citep{shull12}.
This method has the disadvantage that it does not recognize broad absorption features and 
cannot constrain physical properties of the emitting gas from different regions.
In this work, we want to disentangle the radiation from different regions
so we can apply absorption models accordingly,
especially when the intrinsic absorbers in Mrk 290 are likely located between the BLR and the NLR (Z11).
In addition, two adjacent intrinsic \civ~absorption lines ($\sim$1595 \AA) in the COS spectrum are just at the peak of a \civ~emission line, which contaminates the spline fits.
As a consequence, we adopt a different approach to fit the spectrum.  
We first analyze the radiation from the BLR and the NLR and construct XSTAR table models.
We then combine these with a model for the radiation from the accretion disk of Mrk 290, a featureless continuum in the COS band (1140 - 1800 \AA), which we describe as a power law \citep{kriss11}.

The photoionized clouds in the BLR and the NLR show many emission lines, including Ly$\alpha$, \civ, \nv, \siiv, \oiv, \niv, \heii~and other weak lines (Fig.~\ref{fig:cos}).
The profiles of these lines are complex, and this complexity mainly originates from the BLR.
The motions of the BLR clouds might be a superposition of different components, such as Doppler motions, turbulence, shock components, in/outflow components, and rotation \citep[][and references therein]{kollatschny13}, and can be altered by radiation pressure \citep{netzer10}.
Moreover, different geometries of the BLR can conspire with these components to result in similar line profiles.
There is no way to uniquely infer the global BLR motion and geometry from line profile fitting.
Line broadening is affected by cloud motion which can be a combination of rotational and radial.    
A plausible assumption is that the cloud speeds are approximately virial, i.e. $v\propto \sqrt{{\rm G}M/R}$ where $M$ is the central mass, 
and $R$ is the distance from the center.  If so, 
the BLR can consist of several groups of clouds with different speeds, and faster the speed of the component, the closer it is to the BH.
Each velocity component is assumed to have the same ionization condition and share one FWHM value to address the line broadening.

\begin{figure}  
   \centering
   \includegraphics[angle=0,width=3.5in]{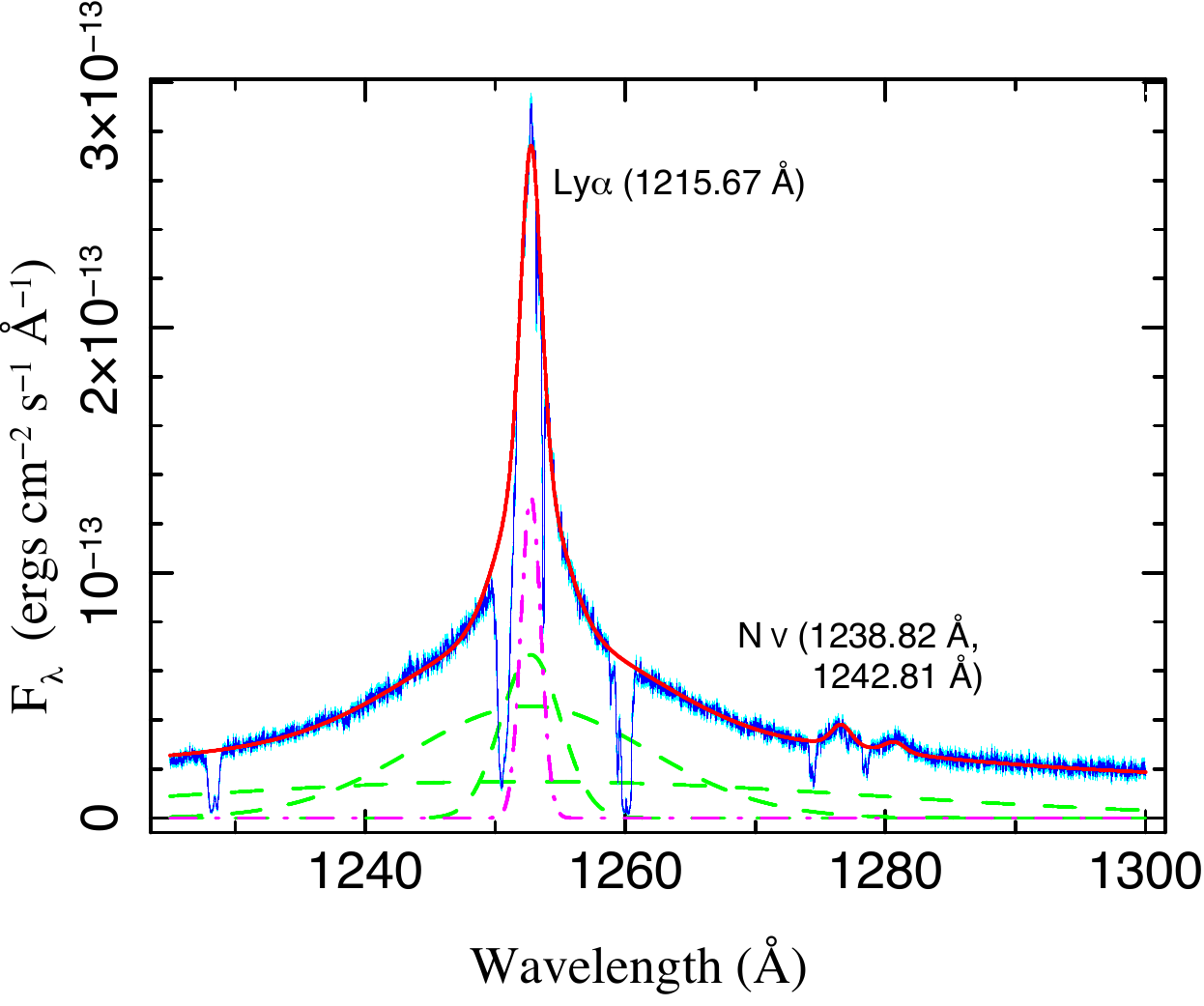} 
   \caption{Ly$\alpha$ and the \nv~doublet are fitted with four groups of Gaussians (red). The dash lines (green) represent the three BLR components that contribute to Ly$\alpha$, while the dash-dotted line (purple) shows the Ly$\alpha$ emission from the NLR.}
   \label{fig:lymana}
\end{figure}

The best target for  profile decomposition is the Ly$\alpha$ emission line combined with the \nv~doublet (1238.8, 1242.8 \AA), which has high signal-to-noise ratio in the COS spectrum.
A summation of several velocity components from both the BLR and the NLR is adopted to mimic their line profiles.
Each velocity component has three Gaussians with the same FWHM to contribute to the three lines, two of which for the \nv~doublet have an optically thin 2:1 flux ratio between the strong and weak lines.
The wavelength band of 1225--1300 \AA~is included for the fit, while all the absorption features are removed or fitted with negative Gaussians.
Finally, at least four velocity components are required to give a good fit to the line profiles of the Ly$\alpha$ and the \nv~doublet simultaneously (Fig.~\ref{fig:lymana}).
The FWHMs of the four components are distinctive: 10183$\pm$344, 5550$\pm$35, 1351$\pm$22 and
417$\pm$6 \kmps.
The narrowest one of 417 \kmps~is consisted with that of the forbidden lines [\oiii] (4959, 5007 \AA) in the Sloan Digital Sky Survey (SDSS) spectrum of Mrk 290 with a FWHM of 400 \kmps, suggesting that this narrowest component originates from the NLR. 
Thus, the other three velocity components are attributed to the BLR.
The redshift of the narrowest component is 0.030540, slightly larger than the redshift 0.0304 obtained from the SDSS spectrum that we used in Z11.
Since COS has a better spectral resolution than SDSS, we correct the redshift of Mrk 290 to 0.030540.

We generate three XSTAR table models for the BLR and one table model for the NLR.
For them, $b$ values are used in XSTAR for producing equal FWHMs of line profiles: $b=\rm{FWHM}/2\sqrt{(ln2)}\approx FWHM/1.665$.
The density of the BLR is set to $10^{10}\,\rm{cm^{-3}}$, while that of the NLR is set to $10^3\,\rm{cm^{-3}}$ according to the flux ratio of [\sulii] (6716, 6731 \AA) doublet that is $\sim$1 in the SDSS spectrum.
For both, the column density is set to $10^{23}\,\rm{cm^{-2}}$, the temperature is set to 15000 K, 
the metallicities are set to the solar values, while the ionization parameter and the redshift are left as free parameters.
The ionizing SED of Mrk 290 used in XSTAR is similar to the low-state SED in Z11.
The X-ray spectrum is obtained from \xmm~observations in 2007, which is close to the COS observations in 2009, while other data are from NASA/IPAC Extragalactic Database: radio 4.8 GHz data from Very Large Array (VLA), 25 $\mu$m data from Infrared Astronomical Satellite (IRAS) and r-band data from SDSS.
For the UV data point, we adopt the extinction-corrected flux at 1500 \AA~of the COS spectrum to avoid contamination from emission lines.

The clouds in the BLR partially cover the ionizing source, which means that the BLR emission is from two parts: light transmitted by the clouds in the forward direction to observers, and light in the backward direction reflected by the clouds.
XSTAR calculates spectra for the both parts, which have quite different line ratios.
For the simplicity, we set the parameter of covering factor in XSTAR as 0.1 for the BLR components, and
define an additional parameter ($R_{\rm norm}$) to indicate deviations of the covering fraction to the fixed value of 0.1.
This free parameter is the ratio of `normalization' values of the two spectra for the transmitted and reflected emission 
(the other parameters in the XSTAR model are the same for the two spectra).
The BLR covering fraction will become larger when $R_{\rm norm}$ increases,
because more clouds are blocking the reflected emission along our line of sight.

\subsection{Absorption by the Ionized Outflow}
\begin{figure*}  
   \centering
   \includegraphics[angle=0,width=\textwidth]{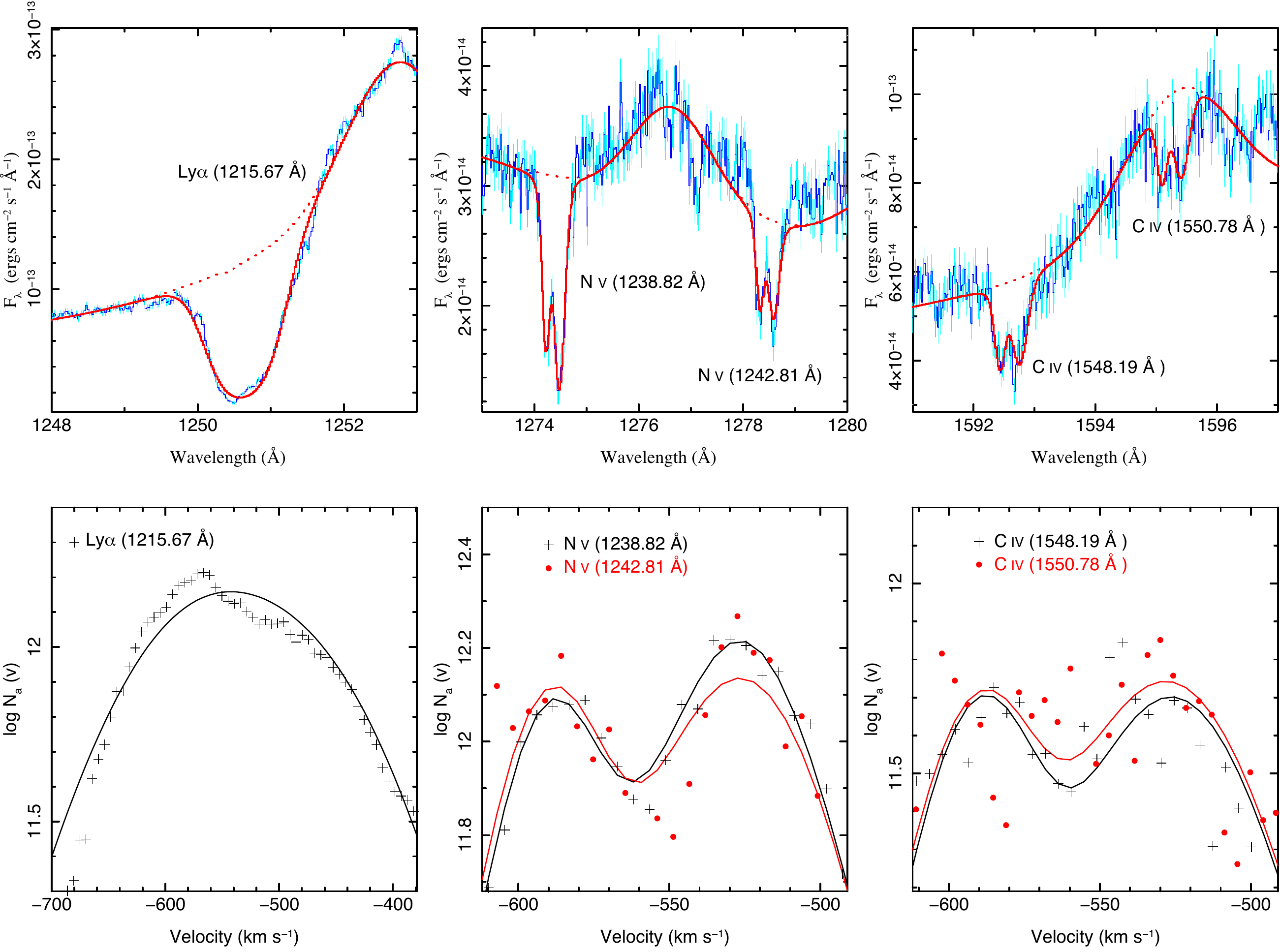} 
   \caption{The upper panels are the close-up of the intrinsic absorption lines. The solid red curve is the best fit model with negative Gaussians for the absorption lines, while the dash curve is the background continuum.
The lower panels are the apparent $N_a$ per unit velocity for the Ly$\alpha$ line and the \nv~and \civ~doublets.
The plus (black) and dot (red) points are derived from the data spectrum, while the solid lines are obtained from the best fit model.}
   \label{fig:na}
\end{figure*}

In the COS spectrum of Mrk 290, Ly$\alpha$, \nv~and \civ~doublets are detected near redshift 0.0287.
The \nv~and \civ~lines show clear double troughs (Fig.~\ref{fig:na}), suggesting at least two absorbers.
For the purpose of obtaining turbulent velocities of the absorbers, we fit these nine intrinsic absorption lines with negative Gaussians (Table~\ref{tab:lines}), while the modeling of the background continuum is described in detail in Section 5.1.
When fitting the \nv~and \civ~doublets, we assume that they come from two absorbers,
and thus for each absorber we relate their wavelengths according to the laboratory values and set their FWHMs to be the same.
The blueshift velocities of the two absorbers are -590 and -525 \kmps, and the corresponding FWHMs are 31 and 51 \kmps, respectively.

\begin{figure}  
   \centering
   \includegraphics[angle=0,width=3.3in]{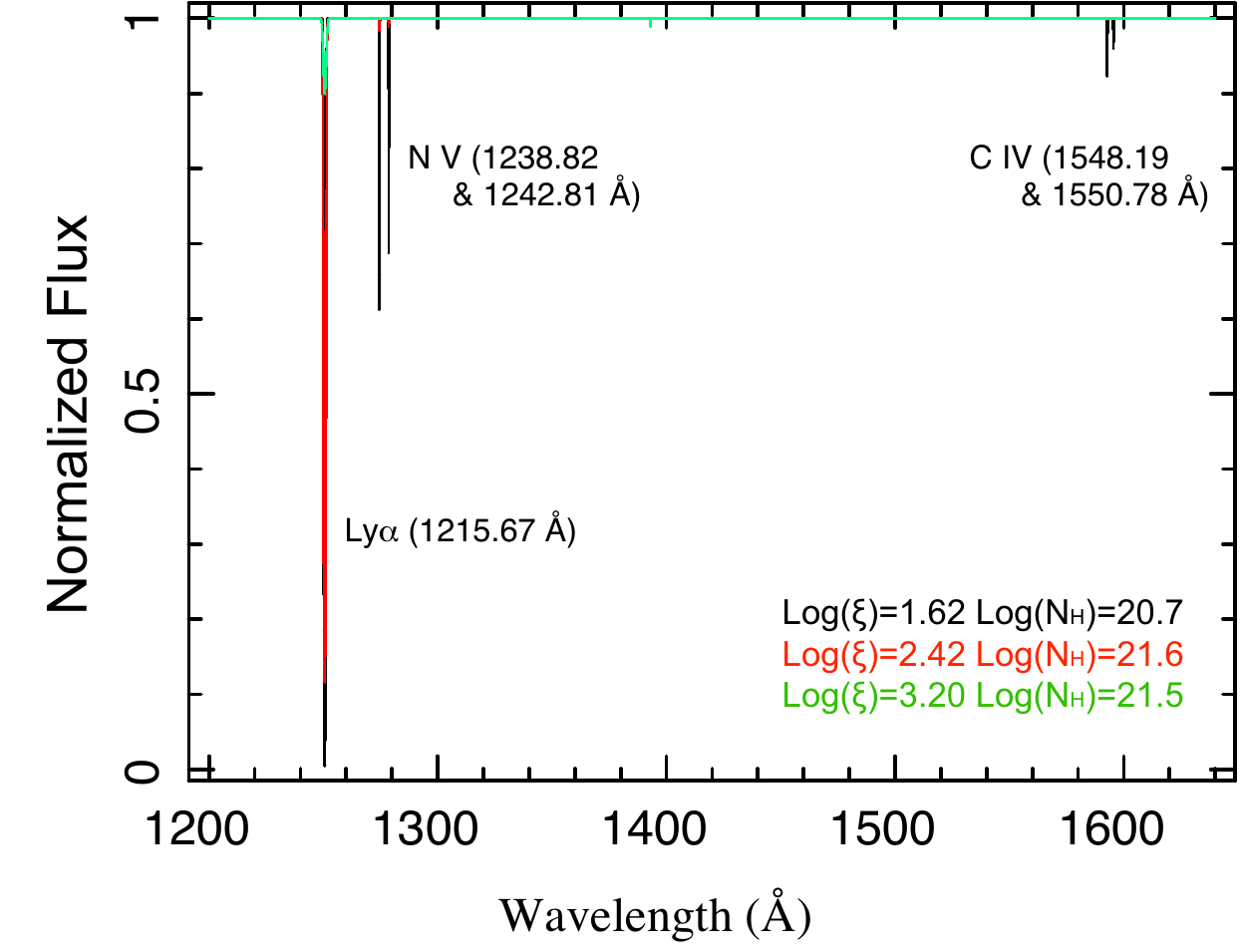} 
   \caption{The predicted UV absorption lines based upon the parameters obtained in the X-ray band from the \chandra~observations.}
   \label{fig:xwa}
\end{figure}

The outflow velocities of the intrinsic UV absorbers are coincident with the WAs measured from the \chandra~X-ray spectra in Z11 (Table~\ref{tab:parameters}).
Given the parameters of the three WAs obtained in the X-ray band, we use XSTAR to predict their behaviors in the UV band (Fig.~\ref{fig:xwa}) and to check the possibility that the intrinsic absorption lines in the two bands are caused by the same ionized gas.
The highest ionization (HI) WA with log$(\xi)=3.2$ does not show any significant absorption, the moderately ionization (MI) WA with log$(\xi)=2.4$ absorbs Ly$\alpha$, and the lowest ionization (LI) WA with log$(\xi)=1.6$ shows Ly$\alpha$, \nv~and \civ~absorption lines.
Thus, the two UV absorbers identified from the double troughs of the \nv~and \civ~doublets are similar to the LI WA, which is detected as one in the X-ray band because of the lower spectral resolution.
The Ly$\alpha$ absorption line may still be affected by the MI WA, especially when the Ly$\alpha$ line is heavily saturated.

A clear smooth truncation at the bottom of the Ly$\alpha$ absorption line (Fig.~\ref{fig:na}) suggests heavy saturation.
It is important to check whether the \nv~and \civ~doublets are saturated, which affects the determination of the ionization states of the absorbers.
We measure the apparent column density per unit velocity interval for each line.
This $N_a(v)$ in logarithmic form is:
$${\rm log}\,[N_a(v)]={\rm log}\,\tau_a(v)-{\rm log}\,(f\lambda)+14.576\;\;\rm{[atoms\,cm^{-2}(km\,s^{-1})]},$$
where the $f$ is the transition oscillator strength (Table~\ref{tab:lines}), the $\lambda$ is the wavelength, and the $\tau_a(v)={\rm ln}[I_o(v)/I(v)]$ is the optical depth that $I_o(v)$ and $I(v)$ are the intensities before and after the absorption (Savage \& Sembach 1991).
A factor of two difference in $f\lambda$, applicable to doublet lines, is often adequate to apply this technique.
When the profiles of the $N_a(v)$ of the strong and weak lines in a doublet are in good agreement,
it is unlikely that unresolved saturated structure strongly influences the observed lines.
The profiles of the $N_a(v)$ of all absorption lines from both the data spectra ($I_{\rm data}(v)$) and the best-fit models with negative Gaussians ($I_{\rm model}(v)$) are shown in Fig.~\ref{fig:na}.
For the \nv~or \civ~doublets, the profiles from $I_{\rm data}(v)$ are significant overlap with each other,
but the scatters are large, especially for the \civ~doublet.
The profiles from $I_{\rm model}(v)$ show discrepancies at -525 \kmps, suggesting some saturation of that absorber.

\begin{deluxetable}{lc|cccc}
\tabletypesize{\footnotesize}
\tablecolumns{6} 
\tablewidth{0pt} 
\tablecaption{Intrinsic Absorption Lines in Mrk 290}
\tablehead{
           \colhead{Ion ($\lambda_{\rm rest}$)}                   &
           \colhead{$f$\tablenotemark{a}}                     &
 	   \colhead{$\lambda_{\rm obs}$}                           &
          \colhead{Flux ($10^{-4}$)}                             &
           \colhead{FWHM}                                          &
           \colhead{$v_{\rm off}$}                               \\
           \colhead{(\AA)}                   &
           \colhead{}                     &
 	   \colhead{\AA}                           &
           \colhead{$\rm{photon\,s^{-1}cm^{-2}}$}                             &
           \colhead{\kmps}                                          &
           \colhead{\kmps}                             
                             }  
\startdata
Ly$\alpha$ (1215.67)   &4.1617e-1 &1250.64$\pm$0.01 & 105.77$\pm$0.69 & 	279$\pm$2 &	-532$\pm$1 	\\
\nv~(1238.82)		   &1.5553e-1 &1274.22$\pm$0.01 &   2.17$\pm$0.16 &	31$\pm$4   &	-590$\pm$2 	\\
                                    &                  &1274.49$\pm$0.01 &   3.83$\pm$0.20 &	51$\pm$7   &	-525$\pm$1 	\\
\nv~(1242.81)	& 7.7805e-2  &  1278.32\tablenotemark{b} & 1.30$\pm$0.15 &   31\tablenotemark{b} &  -590\tablenotemark{b} \\
                         &                    &  1278.59\tablenotemark{b} & 1.98$\pm$0.17 &   51\tablenotemark{b} &	 -525\tablenotemark{b} \\
\civ~(1548.19)  & 1.9045e-1   &  1592.44\tablenotemark{b} & 4.56$\pm$0.42 &   31\tablenotemark{b} &  -590\tablenotemark{b} \\
                         &                    &  1592.77\tablenotemark{b} & 5.24$\pm$0.65 &   51\tablenotemark{b} &	 -525\tablenotemark{b} \\
\civ~(1550.78)  & 9.4824e-2   &  1595.09\tablenotemark{b}  & 6.76$\pm$0.53 &  31\tablenotemark{b} &  -590\tablenotemark{b} \\
                         &                    &  1595.42\tablenotemark{b} & 8.33$\pm$0.76 &   51\tablenotemark{b} & 	 -525\tablenotemark{b} \\
\enddata

\tablenotetext{a}{The transition oscillator strengths are from AtomDB (version 2.0.2; www.atomdb.org).}
\tablenotetext{b}{The values are fixed relative to \nv~(1238.82 \AA).}
\label{tab:lines}
\end{deluxetable}

In spite of the fact that the Ly$\alpha$ absorption line is heavily saturated, the flux at the bottom of the absorption trough is non-zero.
This is partly due to the broad wings on the LSF that are produced by 
the zonal (polishing) errors on the HST primary and secondary mirrors \citep{ghavamian09},
but the dominant reason is likely that the covering fraction of the absorbers is less than 100\%.
Since the BLR is extended, and the continuum may be from an extended region too,
there is a fraction of all photons $C_{\rm f}$ that pass through the absorbing gas \citep{ganguly99}.
If the continuum comes only from the accretion disk, the size is likely significantly smaller than the BLR.
We consider the simplest scenario first, in which 
the continuum emission $I_{\rm cont}$ is fully covered by the absorbing gas,
the BLR emission $I_{\rm BLR}$ is partially covered, while the NLR emission $I_{\rm NLR}$ is not covered at all given that the absorbing gas is torus wind as we argued in Z11.
Then the observed flux after the correction of dust extinction can be expressed as:
$I'_{\rm obs}=(I_{\rm cont}+C_{\rm f}I_{\rm BLR})e^{-\tau}+I_{\rm NLR}$,
where $\tau$ is the corresponding line optical depth.
The total covering fraction to the BLR at each velocity can be obtained directly from the heavily saturated Ly$\alpha$, where the $C_{\rm f}$ varies between 70--80\% within the range of the blueshift velocity between -600 and -450 \kmps.
However, when we test this scenario with the \nv~doublet which has a 2:1 oscillator strength ($f$) ratio between the strong and weak line (note that $\tau\propto f\lambda$; \citealt{hamann97}), 
we find $C_{\rm f}<0$ which is unphysical. 
A possible reason is that the continuum photons are scattered back into the light path again by the clouds in the BLR \citep{ganguly99}.
As a consequence, we consider the covering fraction $C_{\rm f}$ for the continuum source is less than unity, and set it the same as for the BLR to simplify our model and prevent additional free parameters:
$I'_{\rm obs}=(I_{\rm cont}+I_{\rm BLR})C_{\rm f}e^{-\tau}+I_{\rm NLR}$.
In this scenario, the total $C_{\rm f}$ value at each velocity obtained from the Ly$\alpha$ is slightly higher (85--95\%) in the velocity range of -600 -- -450 \kmps.
The covering fraction of each individual absorber will be obtained directly from the modeling.

We produce XSTAR table models for these intrinsic absorbers.
The $b$ values of the two absorbers are their turbulent velocities,  
which can be estimated from the FWHMs of 31 and 51 \kmps~(Table~\ref{tab:lines}).
The FWHM roughly equals $2\sqrt{{\rm ln}2}\sqrt{v_{\rm turb}^2+v_{\rm th}^2}$, where $v_{\rm turb}$ and $v_{\rm th}$ are the turbulent velocity and the thermal velocity, respectively.
When the FWHM values are small, the thermal broadening may play an important role.
The thermal velocity is determined by $v_{\rm th}=13\sqrt{T_4/A}$ \kmps,
where $T_4$ is the temperature in units of $10^4$ K and $A$ is the atomic number.
According to the thermal curve in Z11, the temperature of the absorbers at this ionization state is about 60,000 K, which corresponds to a thermal velocity of $v_{\rm th}\simeq9$ \kmps.
As a result, the turbulent velocities are 16 and 29 \kmps.
The SED is the same as was used for the BLR and NLR. 
The column density, the ionization parameter and the redshift are the free parameters.

\section{Spectral Fitting and Results}

In the COS band, the absorption of the continuum emission by the intrinsic absorbers is negligible,
as also demonstrated in Fig.~\ref{fig:xwa}.
Thus, we can fit the entire COS spectrum of Mrk 290 by two steps.
First, ignoring the intrinsic absorption lines, we fit the AGN emission and the local absorption simultaneously, including the continuum from the accretion disk, the emission from the BLR and NLR, and the absorption from the IGM, HVC, and ISM.
Then, we focus on fitting the intrinsic absorption from the AGN outflows.

\begin{figure*}  
   \centering
   \includegraphics[angle=0,width=\textwidth]{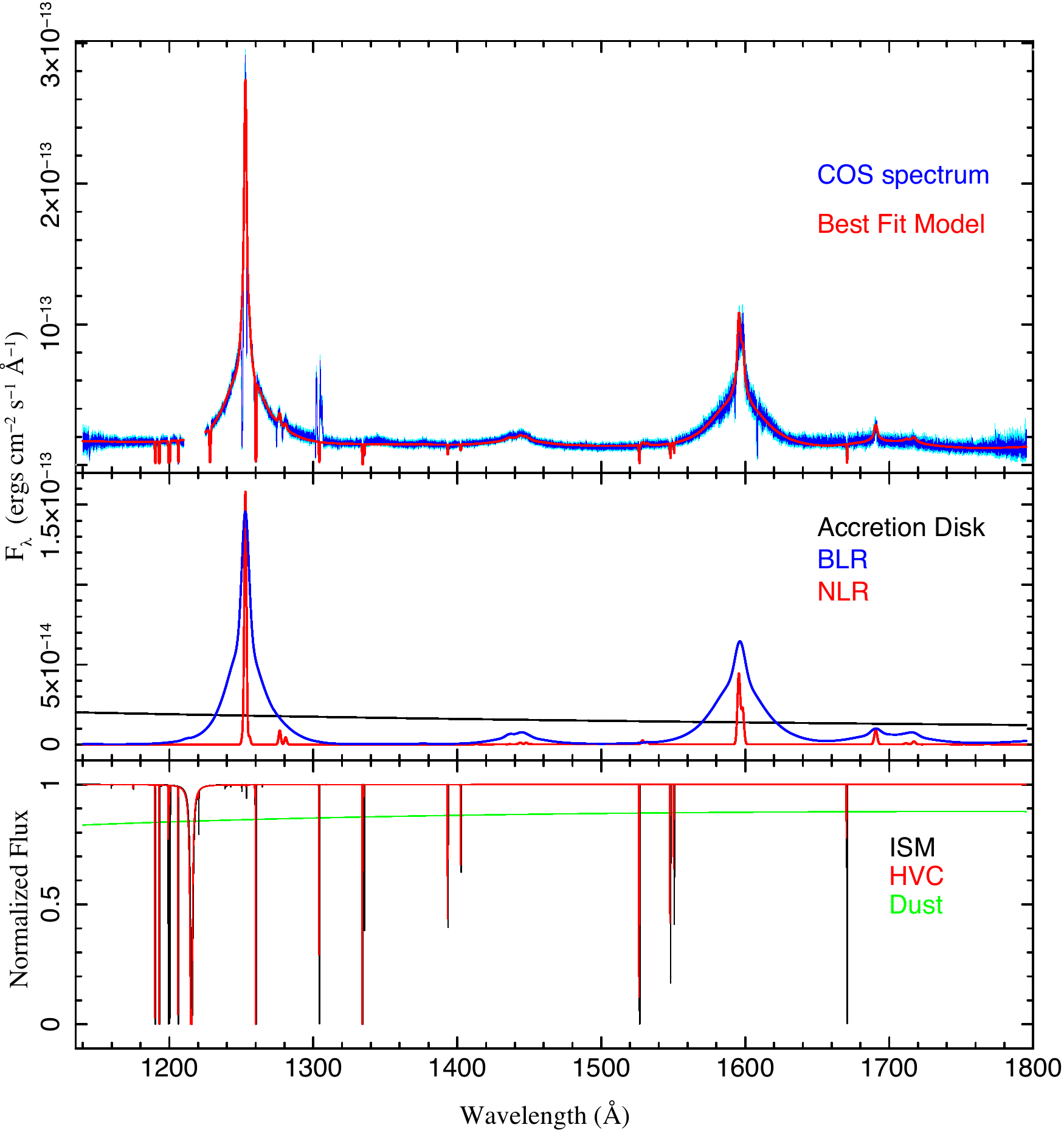} 
   \caption{The top panel represents the best fit model of the AGN emission and the local absorption. The middle panel shows the separate models of the AGN emission components including the accretion disk (black), the BLR (blue), and the NLR (red),
and the bottom panel shows the models of the local absorbtion components including the ISM (black), the HVC (red), and the dust extinction (green).}
   \label{fig:arla}
\end{figure*}

\subsection{AGN emission and local absorption}
 
We fit the AGN emission and the local absorption in the wavelength range of 1140--1795 \AA, 
plus one Gaussian absorption model (`gabs') to account for the broad intrinsic Ly$\alpha$ absorption line.
The geo-coronal features and the intrinsic \nv~and \civ~absorption lines are excluded,
and the HVC component blueshifted at -220 \kmps~is not modeled, which has only a few shallow absorption lines \citep{shull11}.

The fit to the entire COS spectrum is shown in Fig.~\ref{fig:arla}; the model and the data agree within the statistical 
errors over essentially the entire spectral range ($\chi^2$/d.o.f. is 2.1[60090/28342]).
The detailed models of the emitting and absorbing components are also displayed in the figure,
and the parameters of the best-fit photoionization models 
are included in Table~\ref{tab:parameters}.

The combination of the BLR and NLR models reproducs the complex profiles of most emission lines, including Ly$\alpha$, \nv, \niv, \civ, blend of \siiv+\oiv, \heii~and \oiii.
In particular, the models account for the peak of the \civ~emission line that has two absorption lines superimposed (Fig.~\ref{fig:na}).
The ionization parameters of the three BLR components are slightly different but all are at low ionization states.
The $R_{\rm norm}$ of the broadest BLR component is larger than the other two, which suggests a higher covering fraction for the clouds close to the BH.
The continuum from the accretion disk is fitted by a power law with a photon index of $\sim$1.9.

\begin{figure*}  
   \centering
   \includegraphics[angle=0,width=\textwidth]{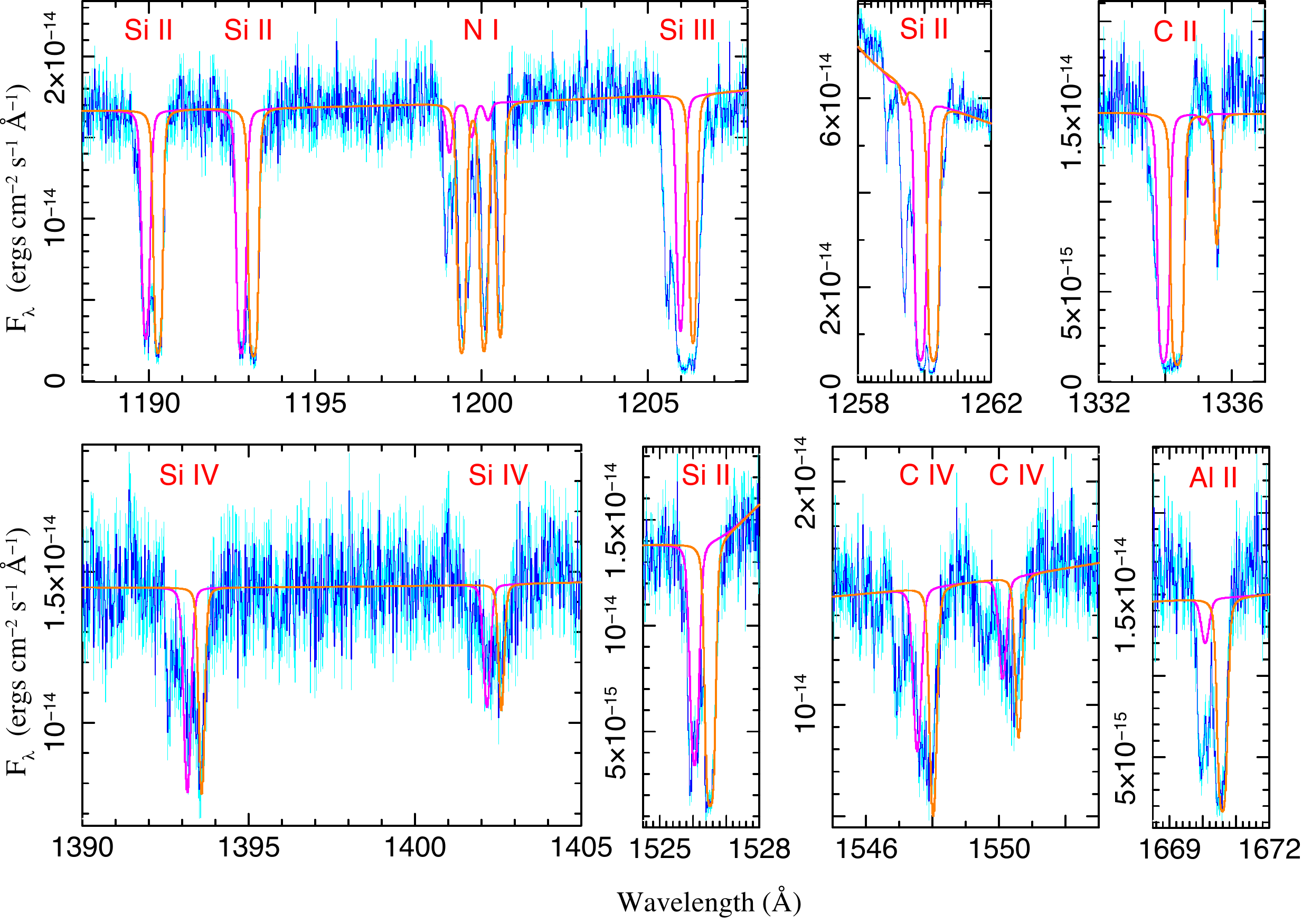} 
   \caption{The absorption models for the HVC (purple) and the ISM warm-hot gas (orange), respectively.}
   \label{fig:ism}
\end{figure*}

Given only one or two free parameters (Table~\ref{tab:parameters}),
the XSTAR table models for the HVC and the ISM warm-hot gas fit well most of the local absorption lines as shown in Fig.~\ref{fig:ism}.
The lines with significant differences between the HVC and the ISM are nicely represented by the models, such as \ni~(at 1200 \AA) and \cii~(at 1335 \AA).
These differences are mainly due to the lower N and C abundances and the higher ionization parameter in the HVC than in the ISM.
However, the observed \ni~triplet of the HVC absorption is deeper than the predicted model, 
suggesting that the abundance of N may be higher than the input value (1.5\% of solar) obtained by \citet{shull11}.
The \alii~line also indicates a higher abundance than the input Al abundance of 4\%.
The authors claimed that ionization corrections were not made for the \ni~and \alii~lines,
so that the N and Al abundances may be less accurate compared to the other elements.
The \siiii~line is not perfectly matched by the models, indicating that the distribution of the ionization states could be more complicated in the HVC and ISM.
The \feii~(1608 \AA) lines and several \sulii~lines upon the broad Ly$\alpha$ emission line
are absent in the local absorption models because of incomplete atomic data of \feii~ and \sulii.

\begin{deluxetable}{lcccc|c}
\tablecolumns{5} 
\tablewidth{0pt} 
\tablecaption{The best fit parameters of photoionization models}
\tablehead{
           \colhead{ }                &
           \colhead{Log($\xi$)}          &
           \colhead{$N_{\rm H}$}            &
           \colhead{redshift}        &
           \colhead{velocity\tablenotemark{*}}        &
           \colhead{ }  \\
           \colhead{ }                                                    &
	  \colhead{($\rm{erg\,s\,cm^{-1}}$)}                     &
           \colhead{($\rm{10^{20}\,cm^{-2}}$)}             &
           \colhead{ }                                                    &
           \colhead{(\kmps)}              &
           \colhead{     }                  
                      }  
\startdata
\multicolumn{5}{l}{Local Absorption:} \\
HVC   & -0.74$\pm$0.04 & 1.12 (fixed) & -0.000430$\pm$0.000001   &  -129.0$\pm$0.3 &  \\
ISM    & -1.29$\pm$0.01  & 0.57$\pm$0.01 & -0.000121$\pm$0.000001   &  -36.3$\pm$0.3 & \\
    \hline\noalign{\smallskip}
    \multicolumn{5}{l}{AGN Radiation:} & $R_{\rm norm}$\tablenotemark{a} \\
BLR {\small I}  & -0.601$\pm$0.007  & - &  0.030546$\pm$0.000013   &     1.8$\pm$3.9   & 5.0$\pm$0.4  \\
BLR {\small II}  &  0.659$\pm$0.004  & - &  0.029855$\pm$0.000035  & -205.5$\pm$10.5 & 1.2$\pm$0.1 \\
BLR {\small III}  &  0.125$\pm$0.009  & - &  0.029872$\pm$0.000052  & -200.3$\pm$15.6 & 20.7$\pm$1.9 \\
NLR       &  0.592$\pm$0.028  & - &  0.030540$\pm$0.000005  &         0$\pm$1.5   &- \\
    \hline\noalign{\smallskip}
\multicolumn{5}{l}{Intrinsic Outflow Absorption:}  & $C_{\rm f}$\tablenotemark{b}\\
WA {\small I} & $1.13\pm0.10$ & $0.5\pm0.2$ & $0.028570\pm0.000005$   & -591.0$\pm$1.5  & 66$\pm$3(\%) \\
WA {\small II} & $1.39\pm0.04$ & $4.8\pm1.4$ & $0.028789\pm0.000006$   & -525.3$\pm$1.8  &  56$\pm$3(\%) \\
WA {\small III} &  2.2 (fixed)     & $36.0\pm0.8$  & $0.028811\pm0.000005$   & -518.7$\pm$1.5 & 72$\pm$3(\%) \\
    \hline\noalign{\smallskip}
\multicolumn{6}{l}{X-ray WAs from \chandra~observations (Z11):}  \\
~LI WA &  $1.62\pm0.15$ & $5.4\pm1.8$  & -   & -570$\pm$150 & - \\
MI WA & $2.42\pm0.04$ & $40.6\pm6.2$ & -   & -480$\pm$30  & - \\
~HI WA & $3.20\pm0.14$ & $35\pm15$ & - & -390$\pm$60  & - \\
\enddata
\tablenotetext{*}{The velocities of the HVC and the ISM are relative to heliocentric coordinates, while other components are relative to AGN rest frame. The parameters of X-ray WAs are listed here for comparison, where the `LI', `MI' and `HI' WAs mean the WAs at the lowest ionization, the moderately ionization, and the highest ionization states.}
\tablenotetext{a}{The $R_{\rm norm}$ is the ratio of `normalization' values of the two spectra for the transmitted and reflected emission by the BLR clouds, which indicates the degree of covering fraction to the BH.}
\tablenotetext{b}{The $C_{\rm f}$ is the covering fraction to the continuum and BLR emission by each absorber.}
\label{tab:parameters}
\end{deluxetable}

\subsection{Ionized Outflow Absorption}
The two photoionization models with the turbulent velocities of 16 and 29 \kmps~are applied to fit the two intrinsic absorbers
identified from the \nv~and \civ~doublets.
However, the Ly$\alpha$ absorption line may be also contributed by absorbers such as the MI X-ray WA (log$\xi=2.42$) as predicted by XSTAR models (Fig.~\ref{fig:xwa}).
One more intrinsic absorber (or WA) is added to account for the extra absorption of Ly$\alpha$
by applying a Gaussian absorption model (`gabs').
Finally, three intrinsic UV absorbers are modeled in the COS spectrum,
and each of them couples with a covering fraction $C_{\rm f}$ as discussed in Section 4.3.

\begin{figure*}  
   \centering
   \includegraphics[angle=0,width=6in]{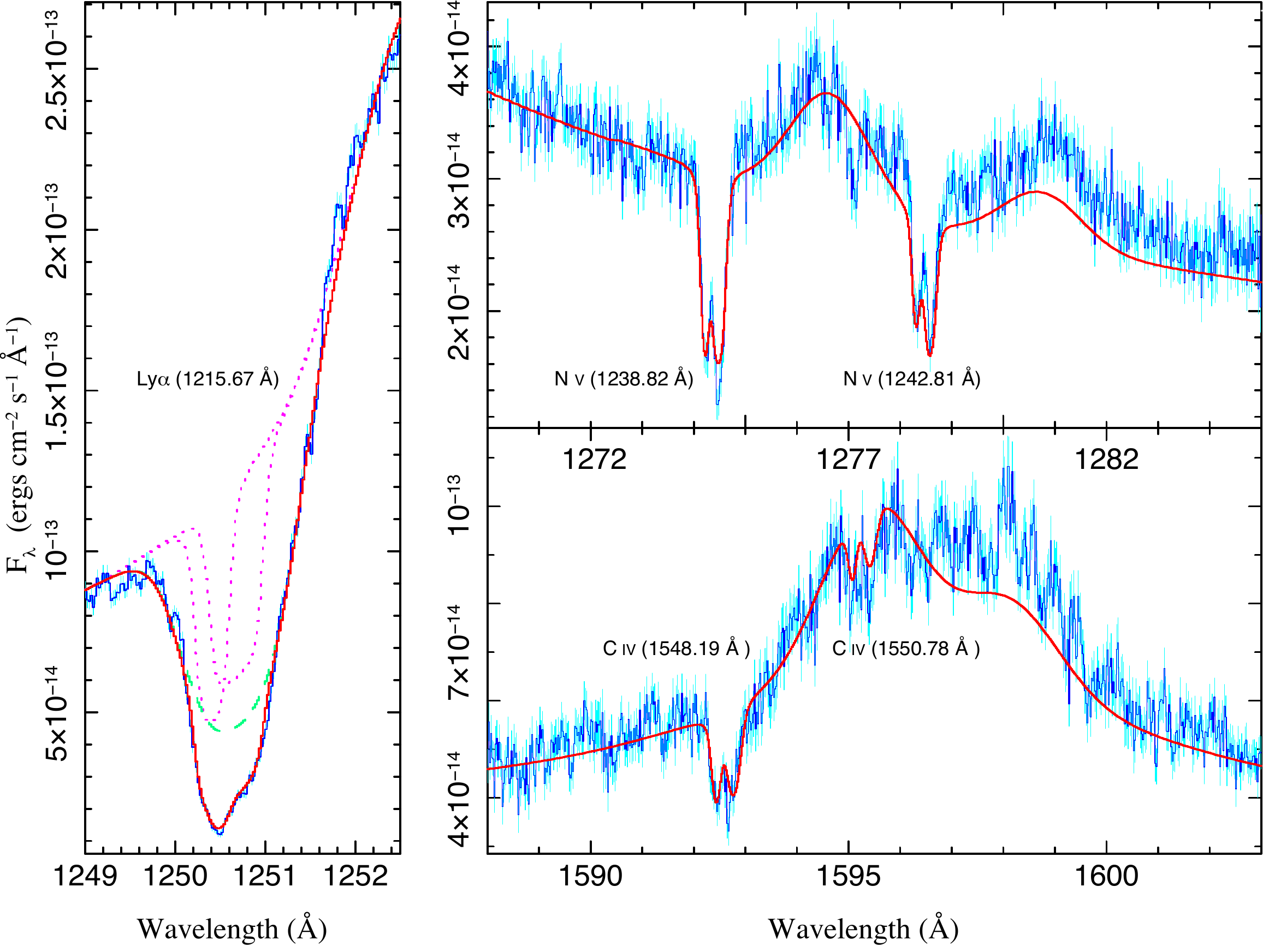} 
   \caption{Three photoionized absorption models give the best fit (red) to all the intrinsic absorption lines. The dotted lines (purple) in the left panel are contribution from WA {\small I} and {\small II} which also absorb the \nv~and \civ~lines. The WA {\small III} only affects Ly$\alpha$ as shown with the dash line (green).}
   \label{fig:wacos}
\end{figure*}

The two photoionization models fit the \nv~and \civ~doublets, and with the additional `gabs' model
they can represent the complicated shape of the Ly$\alpha$ line (Fig.~\ref{fig:wacos}).
The best fit parameters are listed in Table~\ref{tab:parameters}, where the two absorbers have similar ionization states (log$\xi$=1.13 \& 1.39) and covering fractions (66\% \& 56\%).
However, the second absorber with an outflow velocity of -525 \kmps
(referred to as absorber {\small II}) has a column density
ten times higher than the absorber with a velocity of -591 \kmps
(referred to as absorber {\small I}).
The `gabs' model constrains the outflow velocity of the third absorber (absorber {\small III})
as -519 \kmps, the FWHM as 195$\pm$5 \kmps, and a higher covering fraction as 72\%.

\section{Discussion}

\subsection{One-to-one Correspondence of WAs}

Based on the ability to model the broad band spectrum using XSTAR,
the physical properties of absorbers obtained from the UV and X-ray bands can be compared directly.
If the absorption lines describe the same absorbing gas, then
the information combined from the two bands give constraints on the intrinsic absorbers.
For Mrk 290, the COS observations were taken in 2009, 6 years after the \chandra~observations in 2003.
But the three intrinsic absorbers we detected in the COS spectrum are very similar to
the two X-ray WAs at moderate and low ionization states (Table~\ref{tab:parameters}).
The velocities of absorbers {\small I} and {\small II} are separated by 66 \kmps,
well within the velocity error bars of the LI WA.
Since the column density of absorber {\small II} is much larger than {\small I},
its contribution to the X-ray absorption is relatively more important.
The absorber {\small II} has a column density in good agreement with that of the LI WA,
and an ionization parameter only $\sim$0.2 dex lower
which may be due to a decrease in the ionizing luminosity.
Nevertheless, the UV absorbers {\small I} and {\small II} are likely 
the same gas described by the LI WA,
and we call them WA {\small I} and WA {\small II} from now on.
 
We attribute absorber {\small III} to the extra absorption of Ly$\alpha$ 
beside that of WA {\small I} and {\small II}.
As shown in Fig.~\ref{fig:xwa}, the absorber {\small III} should have 
ionization state similar to that of the MI WA,
otherwise it will absorb the \nv~and \civ~lines at lower ionization state, or won't absorb Ly$\alpha$ line at all at higher ionization state.
We replace the `gabs' model with a photoionization model to deduce
a reasonable column density for this component.
According to the variation of log$\xi$ in WA {\small II}, the ionization parameter of absorber {\small III} is fixed at 2.2 which is 0.2 dex lower than that of the MI WA (log$\xi$=2.4).
The turbulent velocity ($b$ value) of 115 \kmps~is obtained from the FWHM, given that the temperature is 10$^5$ K (Z11).
This new XSTAR model results in a column density of $3.6\times10^{21}\,\rm{cm^{-2}}$, in good agreement with the MI WA value of $4.1\times10^{21}\,\rm{cm^{-2}}$ (Table~\ref{tab:parameters}).
As a result, the absorber {\small III} and the MI WA are reasonably the same gas,
so we call it WA {\small III}.

\subsection{Outflow Geometry}
Various WA outflow geometries have been proposed, such as wind from the accretion disk \citep{elvis00} or putative torus \citep{krolik01}, and clouds in BLR \citep{kraemer05} or NLR \citep{behar03}.
Whatever, all the spectroscopic modeling in this work is consistent with the previous argument that
the WAs are the thermal wind from the inner side of the torus (Z11).
The turbulent velocities of the three WAs detected in the COS spectrum support this scenario, 
which are small ($v_{\rm turb}\lesssim100$ \kmps) due to mild thermal evaporation.
For example, if the WAs are launched from the accretion disk, broad resonance UV absorption lines would be expected \citep{proga04}.

The WAs cover more than 70\% of the BLR in the range of the blueshift velocity between -600 and -450 \kmps.
Given the covering fraction of $\sim$65\% for each WA, the lengths of the WAs should be roughly the same order of magnitude as the radius of the BLR in Mrk 290, which is $\simeq$0.007 pc (or $2.16\times10^{11}$ km) constrained by reverberation mapping \citep{denney10}.
Based on the X-ray variations, the densities of WA {\small I} and {\small II} are constrained to be $n_{\rm e}>5.5\times10^4\,\rm{cm^{−3}}$, while that of WA {\small III} is constrained to be $n_{\rm e}<6.2\times10^4\,\rm{cm^{−3}}$ (Z11).
As a result, the thicknesses of WAs {\small I} and {\small II} should be less than $10^{10}$ and $10^{11}$ km respectively, while that of WA {\small III} should be larger than $6\times10^{11}$ km.
Thus, the lengths and thicknesses are comparable, and the geometry of WAs are more likely discrete clouds with radii of $\sim$$10^{11}$ km rather than flat and thin layers of continuous wind.

In the X-ray study, the mass outflow rate contributed by WAs is $\sim0.1\,\rm M_{\odot}\,yr^{-1}$ assuming a bi-conical chimney-like continuous-wind geometry with a total opening angle less than $\pi$ (Z11).
The covering fraction obtained in the UV band implies a value about half as large.
No new WA with distinctive velocity or significant column density is identified, which suggests an even lower mass outflow rate.
However, the time scale for the WAs to move outside the BLR is about 14 years, 
assuming that they have a transverse velocity of $\sim$500 \kmps~similar to the observed radial velocities, which is larger than the escape velocity of $\sim$440 \kmps~at the inner side of the torus (Z11).
This is consistent with the non-detection of new WAs in the 6 year span between the two sets of observations.
This also adds to the plausibility of the one-to-one correspondence of the WAs.
The HI WA in the X-ray band is not detected in the COS spectrum.
But this tenuous, highly ionized material may still surround the lower ionized clouds, 
contributing to half of the mass outflow rate.

\section{Conclusions}

We have modeled the entire HST/COS spectrum of Mrk 290 with photoionization models generated by XSTAR.
All the physical components in the COS spectrum are decomposed, and the properties of the intrinsic absorbers are compared between the UV and X-ray band straightforwardly, because XSTAR has full coverage of the UV and X-ray band.
We summarize the main results and conclusions as follows:

\begin{itemize}
\item With only one or two free parameters for each component, the XSTAR models fit nearly all the local absorption lines from the HVC (the component blueshifted at -120 \kmps) and the ISM. The model for the HVC, however, suggests higher N and Al abundances than the values measured by \citet{shull11}.

\item The AGN emission from the accretion disk, the BLR and the NLR is disentangled,
which allows us to apply absorption models for outflows that may originate from the inner edge of the torus. 
The clouds in the BLR and NLR are all at low ionization states.
The modeling of the BLR emission suggests a higher covering fraction of clouds when they are very closer to the BH.

\item Three intrinsic UV absorbers outflowing with a radial velocity $\sim$500 \kmps~are identified,
which are consistent with the two WAs obtained in the X-ray band.
The WAs are likely in the geometry of clouds rather than flat and thin layers.
Their small turbulent velocities ($v_{\rm turb}\lesssim100$ \kmps) also support the scenario that the WAs are from the torus due to thermal evaporation.

\end{itemize}

\section*{Acknowledgements}
The anonymous referee is thanked for careful reading of the manuscript and for helpful comments.
We thank John Houck for solving all ISIS software problems, and Yanmei Chen for dealing with the SDSS spectrum of Mrk 290.
We are grateful to MIT Kavli institute for supplying computing time of clusters.
The work is partly supported by the National Natural Science Foundation of China under the grant 11203080. 
Li Ji is also supported by the 100 Talents program of Chinese Academy of Sciences.


\end{document}